\documentclass[numbers,preprint, review, 12pt]{elsarticle}

\usepackage{amssymb}
\usepackage{amsmath}
\usepackage{subcaption}
\journal{Journal of Parallel and Distributed Computing}

\begin{document}
\begin{frontmatter}

\title{Improving dynamic congestion isolation in data-center networks}

\author[uclm]{Alberto Merino}
\ead{Alberto.Merino@uclm.es}

\author[uclm]{Jesus Escudero-Sahuquillo}
\ead{Jesus.Escudero@uclm.es}

\author[uclm]{Pedro Javier Garcia}
\ead{pedrojavier.garcia@uclm.es}

\author[uclm]{Francisco J. Quiles}
\ead{francisco.quiles@uclm.es}

\affiliation[uclm]{
    organization={Universidad de Castilla-La Mancha},
    addressline={Altagracia, 50}, 
    city={Ciudad Real},
    postcode={13071}, 
    state={Ciudad Real},
    country={Spain}
    }

\begin{abstract}
The rise of distributed AI and large-scale applications has impacted the communication operations of data-center and Supercomputer interconnection networks, leading to dramatic incast or in-network congestion scenarios and challenging existing congestion control mechanisms, such as injection throttling (e.g., DCQCN) or congestion isolation (CI). While DCQCN provides a scalable traffic rate adjustment for congesting flows at end nodes (which is slow) and CI effectively isolates these flows in special network resources (which requires extra logic in the switches), their combined use, although it diminishes their particular drawbacks, leads to false congestion scenarios identification and signaling, excessive throttling, and inefficient network resource utilization. In this paper, we propose a new CI mechanism, called Improved Congestion Isolation (ICI), which efficiently combines CI and DCQCN so that the information of the isolated congesting flows is used to guide the ECN marking performed by DCQCN in a way that victim flows do not end up being marked. This coordination reduces false-positive congestion detection, suppresses unnecessary closed-loop feedback (i.e., wrong congestion notifications), and improves responsiveness to communication microbursts. Evaluated under diverse traffic patterns, including incast and Data-center workloads, ICI reduces the number of generated BECNs by up to 32x and improves tail latency by up to 31\%, while maintaining high throughput and scalability.
\end{abstract}

\begin{keyword}
Congestion Control \sep HPC \sep Data Center Network \sep Congestion Isolation  
\end{keyword}
\end{frontmatter}

\section{Introduction}
\label{sec:intro}
Recent advances in artificial intelligence (AI), particularly in the development and deployment of large language models (LLMs), have significantly increased the computational and communication demands \cite{mehonic_brain-inspired_2022} of modern computing infrastructures, such as Supercomputers (SCs) and Data centers (DCs). As a result, these systems are experiencing unprecedented demands from applications to deliver higher throughput, lower latency, and improved scalability. These infrastructures typically comprise thousands of computing and memory end nodes interconnected via high-speed networks. The design of these interconnection networks is a critical factor in the overall performance of SCs and DCs, as these networks must satisfy the stringent communication requirements of parallel and distributed applications, otherwise becoming the system bottleneck.

The effectiveness of high-performance interconnection networks depends heavily on their ability to handle millions of communication operations with minimal delay and data loss. Indeed, the interconnection network must not only support high bandwidth but also ensure a minimal tail latency, especially for latency-sensitive tasks such as synchronous operations, barrier synchronization, or remote memory accesses. To achieve this, the interconnection network must be carefully designed, considering the physical topology, routing algorithm, buffering organization at switches, flow control, and congestion control mechanisms. However, as the described applications are becoming increasingly communication-intensive, even the most advanced interconnection network designs are threatened by network congestion and its dramatic effects.

Network congestion occurs when the traffic load exceeds the available capacity of a network link or switch buffer, typically due to bursty or synchronized traffic patterns such as incast. When this happens at a switch egress port, backpressure is propagated upstream through link-level flow control mechanisms like Priority-based Flow Control (PFC), which pauses packet transmission from upstream neighbors to avoid packet loss. This backpressure propagation creates a congestion tree: a directed subgraph at the bottleneck source (i.e., the congestion root) that includes all upstream switches and hosts (i.e., the congestion branches) that address traffic to that congestion root. Although PFC ensures lossless communication, which is critical, for instance, in RDMA-based systems, the PFC backpressure also produces Head-of-Line (HoL) blocking \cite{garcia19nendica}, which is considered the main negative effect of congestion. Specifically, HoL blocking occurs when non-congesting (“victim”) flows share the same buffer space at switches with congesting flows, the former being unnecessarily stalled, even though they are not addressed to the congestion root. Flow-control backpressure propagates HoL blocking from the congestion root backwards across multiple hops, increasing queuing delays, reducing network throughput, and throttling end-host injection rates. Additional detrimental consequences include unfair buffer monopolization by aggressive flows (buffer hogging)\cite{4289718} and significant degradation of tail latency.

Over the years, diverse strategies have been proposed and implemented in commercial interconnection networks (e.g., InfiniBand or Ethernet) to mitigate congestion and its side effects. Reactive techniques, such as Injection Throttling (ITh), dynamically limit the injection rate at end-hosts of congesting packets based on a closed-loop signaling from the destination servers that receive marked congesting packets, which are first marked at switches whenever congestion is detected at them. 
Examples of ITh techniques are Data Center Quantized Congestion Notification (DCQCN) \cite{dcqcn}, used in Ethernet-based networks (e.g., RoCE) or InfiniBand congestion control (IB-CC) \cite{1208949}.
Adaptive routing has also been used to mitigate the effects of congestion. Still, it may spread the congestion backward through the multiple routes used to route packets. Hence, it must be used carefully and combined with mechanisms that restrict adaptivity \cite{ROCHERGONZALEZ202146} and distinguish between in-network and incast congestion \cite{ROCHERGONZALEZ202427}.
The main drawback of previous techniques is that they do not completely avoid HoL blocking, which is essential to guarantee the network performance under eventual, heavy, and dramatic congestion situations \cite{duato05recn}.

In parallel to the evolution of the previous approaches, other techniques have been devised to reduce or eliminate the HoL blocking specifically.
The most successful ones are the Congestion Isolation (CI) techniques, such as RECN \cite{duato05recn} or DRBCM \cite{6336746}, aimed at eliminating HoL blocking by dynamically identifying the congestion root (i.e., the egress port where buffer occupancy exceeds a threshold) and isolating the congesting flows into dedicated Congested Flow Queues (CFQs) at switch buffers. This per-flow segregation ensures that non-congesting or victim flows remain unaffected, preserving low latency and high throughput under congestion scenarios. However, CI-based techniques require special control logic at switches to keep track of congestion roots, which increases the switch complexity. Preserving this logic as straightforward as possible is a challenge for CI-based techniques, which suffer scalability limitations due to finite hardware resources. Note that the congestion dynamics in the network may generate a high number of concurrent congestion roots whose origin changes according to the specific traffic pattern. To keep track of all these situations, allocating and deallocating resources at switches according to congestion evolution, makes CI control logic even more complex and expensive \cite{6336746}. Another shortcoming of CI mechanisms is that they do not guarantee a fair scheduling among congesting flows, contributing to the same congestion root, thus increasing the tail latency of certain congesting traffic flows.

Despite the individual efficacy of the proposed congestion control techniques, combining multiple congestion control techniques within a single system often introduces new challenges. These approaches typically operate at different layers of the protocol stack and follow divergent design philosophies. As a result, they may not interact harmoniously when deployed together, leading to inconsistent or suboptimal behavior. For example, combining flow control and congestion notification can introduce complex feedback loops, while isolating traffic in virtual lanes may interfere with load balancing or fairness objectives. Furthermore, implementing hybrid approaches requires additional hardware support, firmware customization, or software abstraction layers. These elements increase the design complexity and may introduce new sources of latency and performance overhead.
This challenge becomes even more pronounced when congestion control mechanisms are combined without adequate coordination. Rather than complementing each other, these mechanisms may interfere, leading to counterproductive outcomes that degrade performance more than if used independently. The lack of communication between layers or components within the network stack exacerbates congestion-related issues, particularly under high load or adversarial traffic conditions.

To mitigate these drawbacks, this paper explores the synergistic integration and refinement of existing congestion control mechanisms, particularly DCQCN and CI, in interconnection networks of SCs and DCs, whose architecture converges towards a similar design \cite{Hoefler2022TheCO}. Our goal is to enhance congestion control by leveraging the strengths of each congestion control approach while minimizing their incompatibilities.
In this work, we propose a new technique, called Improved Congestion Isolation (ICI), which enables CI-based mechanisms to provide richer congestion state information to DCQCN-like mechanisms, thereby allowing for a more accurate and responsive congestion signaling to end nodes. Hence, they throttle the injection of congesting flows accordingly. Also, we improve CI-based techniques so they do not generate false positives in congestion detection. We also simplify and reduce the use of network resources devoted to congestion control at switches, allocating and releasing them according to the congestion dynamics.
In summary, the main contributions of this paper are the following:
\begin{itemize}

\item We have designed a set of enhancements to the CI-based techniques (e.g., RECN or DVL) to reduce the resources devoted to identifying congestion roots. These resources will be dynamically used when congestion arises and released when it vanishes, requiring less complexity at switches than state-of-the-art CI proposals.

\item We have combined these enhancements to CI techniques with DCQCN, and developed a set of functionalities so that this integration works smoothly, so that DCQCN uses the CI information to perform a smart ECN marking. This approach reduces the number of packet markings (either wrong or correct). It adjusts the congestion notifications from destination servers to source servers (BECNs), sending them only when required, thus not consuming network bandwidth.

\item  The ICI approach also reduces the number of PFC messages exchanged in the network, thus increasing the network throughput (i.e., reducing backpressure) and reducing the tail latency of traffic flows (either congesting or non-congesting).

\end{itemize}

The rest of the paper is structured as follows. Section \ref{sec:prob} further describes the problem statement related to current congestion control approaches. Section \ref{sec:prop} outlines our ICI proposal. Section \ref{sec:results} shows the experiment results and analyses them. Finally, in Section \ref{sec:conclu}, some conclusions are drawn.

\section{Problem Statement}
\label{sec:prob}

End-to-end congestion control (CC) mechanisms face limitations rooted in their reaction latency and their feedback accuracy when signaling congestion to the servers generating congesting flows. Traditional protocols based on slow-start mechanisms, like TCP and its variants, are poorly suited for short flows because their cautious ramp-up phase often ends before full bandwidth utilization. While more aggressive approaches, such as DCQCN, have been designed to utilize full capacity from the start, they remain vulnerable to short-term congestion bursts. In high-speed data center environments, microsecond-scale incast events can quickly overwhelm switch buffers, triggering ECN marking. The closed-loop control nature of DCQCN, which relies on feedback across multiple Round-Trip Times (RTTs), results in delayed responsiveness. This latency leads to over-marking, conservative rate reductions, and slow recovery, ultimately underutilizing available bandwidth during critical intervals \cite{le2023sfc}.

On the other hand, congestion isolation (CI) mechanisms provide a valuable middle ground by segregating misbehaving flows into special queues, which requires extra logic in the switch control plane. As traffic scales and the number of concurrent flows increases, the number of entries required at flow-matching tables to achieve an efficient isolation of congesting flows becomes unfeasible; thus, the switches end up mixing congested and non-congested traffic in the Congesting Flow Queue (CFQ), undermining the isolation these mechanisms aim to provide.
For example, when deterministic routing is used to balance traffic across switches based on destination addresses, each egress port at those switches may receive traffic flows from multiple input ports, targeting overlapping sets of destinations. In this context, the total number of unique flows reaching a given egress increases with both the number of input ports and the number of destinations mapped to that egress.
Note that this growth in traffic flows poses challenges to congestion detection mechanisms that rely on per-flow tracking, such as CI, which uses a congestion flow table (CFT) at every switch with a limited number of flow entries to isolate and identify several congestion tree roots. However, as the number of congestion tree roots increases, the CFT may run out of entries to store the information of these roots, thus decreasing the effectiveness of CI mechanisms. Indeed, the limited capacity of the isolation tables becomes an important drawback, which hinders the CI efficiency, especially when many entries may be required to capture all potential contributors, as it particularly happens during the early stages of congestion generation and evolution \cite{garcia19nendica}.

If the CFT fills with transient or low-volume flows, it becomes more difficult for CI-based techniques to identify and persistently track the flows contributing to congestion. This can lead to either delayed detection or false positives. The degradation is further exacerbated when speedup mechanisms are applied, as they cause the egress to fill earlier and expose the detector to more flows in a shorter time window.
Also, the lack of space in the CFT for flow isolation per input port due to multiple sources converging on the same egress for the same destinations reduces CI techniques' ability to locate and isolate congesting flows effectively. This highlights the importance of aligning routing strategies, CFT sizing, and detection granularity in large-scale topologies.

The CI allocation mechanism utilizes a tuple-based flow identification. When a flow is flagged as congested, an entry corresponding to its identifying tuple (e.g., source/destination addresses and ports) is created in the CFT. Subsequently, all packets matching this entry tuple at the ingress port are stored in a dedicated CFQ. While effective for immediate isolation of congesting flows, this process is vulnerable in high-density traffic scenarios where false positives are more likely to happen, especially where contention starts at switches close to the end nodes. In this situation, an erroneously identified congesting flow will have its tuple generated into the CFT. Because this isolation state is often mirrored at both ingress and egress ports, a single incorrect prediction can propagate rapidly from the egress to the ingress port at a given switch, leading to unnecessary CFT entries utilization.

Furthermore, the CI CFT entries' de-allocation strategy introduces significant challenges for diagnosing the root cause of persistent congestion. An entry in the CFT, keyed by its flow tuple, contains several fields, including a live packet counter for its presence in the CFQ. De-allocation of the entry is triggered only when two specific conditions are met: i) the packet counter for that flow decrements to zero, and ii) a minimum residency timer expires. This mechanical, state-purging process can lead to contextual decoupling of congestion events. Once a flow's tuple is removed, all historical linkage to the previous congestion event is lost. If that same flow begins to cause congestion again, the system treats it as a new, unrelated incident, misattributing the origin to the local switch. The nature of the de-allocation makes it exceedingly difficult to trace recurring issues back to a persistent upstream bottleneck, thus hindering effective mitigation.

Given these challenges, it is evident that no single technique offers a complete solution. Link-level flow control provides an immediate response but lacks scalability and precision. End-to-end control is scalable but slow to react. Isolation techniques offer targeted relief but are constrained by hardware limitations. This motivates the exploration of hybrid approaches that combine the rapid local responsiveness of link-level mechanisms, the scalability of end-to-end control, and the HoL blocking elimination efficiency of congestion isolation. By carefully integrating their complementary strengths, a more robust and adaptable congestion control architecture can be realized for the demands of modern data centers and HPC workloads.

\section{Improved Congestion Isolation}
\label{sec:prop}
This section outlines Improved Congestion Isolation (ICI), designed to overcome the limitations of existing congestion control mechanisms, such as ICI and DCQCN,  by integrating their strengths into a hybrid architecture capable of finer-grained congestion flow isolation and throttling, fast response to transient congestion and efficient HoL blocking elimination and congesting flows injection throttling.

Our design assumes a shared-buffer switch architecture with internal speedup, which guarantees buffering at the ingress and egress ports of a switch. In this architecture, all input ports share a centralized packet buffer, logically partitioned per port, and each output port has dedicated bandwidth access from the shared memory with a read rate that has a speedup over the write rate in the shared buffer. This switch architecture favors output-port-based congestion detection, as queue occupancy directly moves congestion to the egress side ports.

Nevertheless, the architecture also supports input-side queue monitoring, allowing early detection of incipient congestion when traffic patterns suggest upstream pressure. However, due to the centralized buffering and speedup, sustained congestion typically manifests first at the output stage. Thus, ICI defaults to output-port queue occupancy as the primary signal for triggering congestion responses, while retaining the flexibility to incorporate input-side metrics when beneficial.

\section{Performance Evaluation}
\label{sec:results}
In this section, we evaluate the effectiveness of ICI. First, we describe our network simulation model. Then, we describe the traffic patterns we have used to feed the network simulation. Next, we describe the experiment results and analyze the performance metrics.

\subsection{Network simulation model}

To ensure accuracy, we employ a proprietary, event-driven network simulator written in C++, which models interconnection networks with \textit{cycle-level} and \textit{packet-level} accuracy. The simulator has been validated against real-world InfiniBand deployments and prior established models~\cite{dvl-luis}. Also, it has been widely used in several published papers~\cite{ESCUDEROSAHUQUILLO20141802, dvl-luis}, lending confidence to our findings. It supports detailed modeling of topology, routing policies, switch microarchitecture, NIC behavior, and per-link characteristics, including link bandwidth, propagation delay, or virtual lane (VL) flow-mapping strategies.

In this simulation tool, we have modeled a data center network (DCN) composed of several end nodes connected through a non-blocking or oversubscribed topology using multistage interconnection networks (MINs), which are representative of modern DCNs. Link speed is assumed to be 100~Gbps. The core switching infrastructure consists of shared-memory switches, with each port allocated a guaranteed buffer of 512~KB, ensuring fair and predictable queuing behavior under congestion. The total shared buffer is proportionally distributed across ports, following the practical designs observed in high-end switches~\cite{Bai17} for the given link speed, while maintaining isolation guarantees per egress port.

It is important to note that while the switch employs a shared memory architecture, the presence of internal speedup (1.5$\times$) shifts contention to the egress side, where multiple input flows may simultaneously target the same output port. We enforce a per-egress memory to prevent egress buffer exhaustion from overwhelming the shared memory pool (thereby causing upstream congestion or loss of headroom).
Specifically, we limit the amount of shared buffer that any single egress queue can consume to 50\% of its allocated 512~KB (i.e., 256~KB). This ensures that no single output port monopolizes the buffer, preserving sufficient space for ingress buffering and transient burst absorption (headroom). The remaining 50\% of the per-port allocation is reserved for incoming traffic and safety margin, promoting fair buffer sharing and enhancing resilience to microbursts.

This policy effectively decouples egress congestion from global buffer saturation, enabling cleaner isolation of congestion signals and improving the stability of flow control mechanisms such as PFC. It also aligns with practical implementations in modern switches, where soft limits or virtual channels prevent queue overflow ~\cite{Bai17}.

Network links are serial, full-duplex, and pipelined, operating at 100~Gbps, with a significantly reduced propagation delay of 25~nanoseconds (ns), representative of short-reach optical or electrical interconnects in modern rack-scale deployments. This low-latency assumption also reflects current trends in DCNs connectivity, where physical distances are minimized and signaling efficiency is maximized.

We assume deterministic routing based on the $D$-mod-$K$~\cite{Zahavi2010} algorithm, which evenly distributes traffic across available network paths. Each packet has a fixed MTU of 4~KB, aligning with current configurations in RDMA and high-performance Ethernet environments.

Regarding the switch microarchitecture, we assume a switch speedup of 2, meaning the internal fabric can process packets 2 $\times$ faster than the aggregate line rate of input ports. This architecture causes congestion to manifest primarily at the egress side of the switch due to bursty traffic patterns such as incast.

Flow control is modeled using PFC, which ensures a lossless network. PFC thresholds are calculated based on the \textit{Round-Trip Time (RTT)} in terms of packet units.
Given the shared memory model and per-egress buffer capacity guarantee of 256~KB, we configure PFC thresholds to prevent buffer overflow while maintaining high link utilization. For 4~KB packets, each output buffer can store a maximum of $256\,\text{KB} / 4\,\text{KB} = 64$ packets.

To ensure congestion signals are triggered early enough to prevent packet loss, we set the $PFC_{\text{stop}}$ threshold to 50 packets per ingress port. This leaves a headroom of 14 packets (~56~KB) to absorb in-flight traffic during the propagation delay of the pause frame, effectively eliminating packet drops due to buffer overflow.

In configurations with multiple Virtual Lanes (VLs) active simultaneously, we apply proportional threshold allocation to maintain fairness and prevent any single VL from dominating buffer space. For instance, in a setup with two VLs sharing an egress port, we set $PFC_{\text{stop}} = 22$ and $PFC_{\text{stop}} = 17$, ensuring that the aggregate usage remains below the global ingress cap while allowing asymmetric prioritization if needed. These values are derived from traffic class weights and congestion sensitivity, and are coordinated so that the sum of the individual VL thresholds does not exceed the global 50-packet limit.

Similarly, $PFC_{\text{go}}$ is set slightly below $PFC_{\text{stop}}$ (e.g., 45 packets at the port level, or proportionally lower per VL) to prevent rapid oscillation between pause and resume states. This ensures stable flow control operation under bursty traffic conditions.

Together, these settings guarantee that no queue exceeds its allocated memory and, consequently, no packets are dropped due to buffer overflow, preserving the lossless semantics required for ICI to operate effectively.

\subsection{Network workloads and traffic patterns}
\label{sec:workloads}

To comprehensively evaluate ICI and answer the research questions posed in Section~\ref{sec:results}, we design a diverse set of synthetic and trace-based traffic scenarios. These are specifically chosen to stress-test both ICI and DCQCN under conditions where each individually excels, as well as in challenging hybrid workloads where their limitations become evident. Our goal is to demonstrate that ICI not only maintains the strengths of both mechanisms but also mitigates their weaknesses through intelligent coordination.

The evaluated scenarios are as follows:
\begin{itemize}
\item \textbf{1 Congestion Tree}: A synthetic workload where 75\% of nodes generate uniform random traffic, while the remaining 25\% simultaneously send traffic at line rate to a single shared destination for 1~ms, creating a classic incast congestion tree.

    \item \textbf{4 Congestion Tree}: Similar to the above, but the 25\% of senders split their traffic across four distinct destinations, distributing the load and increasing the number of concurrent congestion points.

    \item \textbf{Google All + 1 Congestion Tree}: We combine 2 million messages following the \texttt{Google\_All} flow size distribution~\cite{Montazeri2018homa} with a single incast congestion tree. The destination of the congestion tree is excluded from the \texttt{Google\_All} destination pool to avoid self-interference, ensuring that application-level flows are not artificially penalized due to their proximity to the congestion point.

    \item \textbf{Google All + 1 Congestion Tree Burst}: Same as above, but the incast is divided into 1 burst of 100,000 packets each, allowing us to assess how well ICI+ handles transient, rapidly absorbed congestion events.

    \item \textbf{Google All + 4 Congestion Tree Burst}: Extension of the previous setup with four Burst of 1 incast tree, increasing control-plane pressure and testing scalability under small but rapidly changing congestion.

\end{itemize}

To emulate realistic and dynamic congestion scenarios, we generate synthetic traffic patterns and extend to more realistic scenarios via the \texttt{VEFTrace-Lib} framework~\cite{Andujar23VEFtraces}, which introduces variable burstiness and non-uniform flow arrival rates. This approach enables us to stress-test ICI under transient and sustained congestion, going beyond uniform random traffic. Also \texttt{VEFTraces-lib} enables testing real scenarios by reproducing real traffic traces.

Note that the combination of a validated simulation framework, scalable topologies, and realistic traffic models enables an evaluation of ICI+ across diverse network conditions. In the following, we analyze the results in light of our research questions.

\subsection{Experiment results with one congestion tree}
\label{sec:1-cong-tree}

\begin{figure*}[!htb]
    \centering
    \begin{subfigure}[b]{\linewidth}
      \centering
        \includegraphics[width=0.6\linewidth]{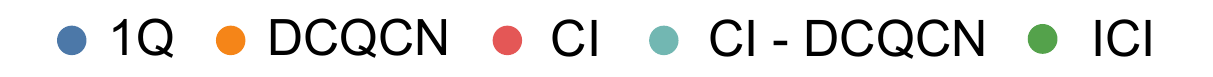}
        \end{subfigure}
        \hfill
      \begin{subfigure}[b]{\linewidth}
      \centering
        \includegraphics[width=\linewidth]{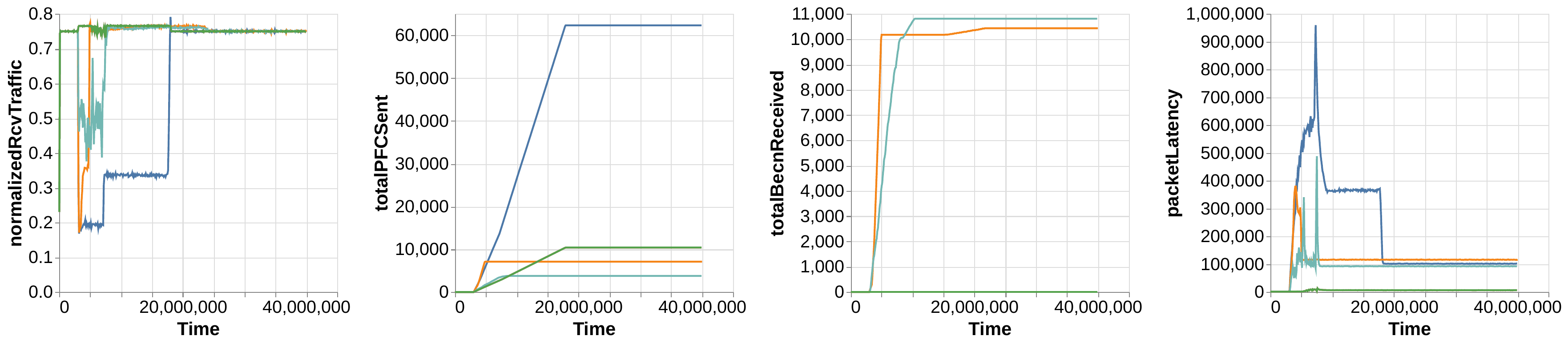}
      \end{subfigure}
    \caption{1HS Traffic pattern 64 MIN network.}
    \label{fig:1-cong-becn}
\end{figure*}

This scenario evaluates system behavior under a single, dominant congestion point, a case where ICI performs optimally due to sufficient capacity in the CFT. Similarly, DCQCN reacts effectively by marking ECN bits early. However, when combining ICI and DCQCN naively (i.e., without coordination), the results in Figure~\ref{fig:1-cong-becn} show a slight performance degradation compared to either mechanism alone.

We attribute this to a \textit{lack of coordination}: both mechanisms act independently, potentially penalizing different subsets of flows, including non-congesting ones. This disjoint reaction leads to suboptimal fairness and unnecessary throttling.

In contrast, ICI+ detects that the congestion is fully manageable by ICI’s flow-state mechanism. As a result, DCQCN is not triggered, and no BECNs or ECN marks are generated (see Figure~\ref{fig:1-cong-becn}). This selective activation demonstrates ICI+’s ability to \textit{assess congestion severity} and avoid redundant control actions.

In terms of latency, both ICI and ICI+ achieve the lowest packet and flow completion times, confirming that ICI+ preserves ICI’s efficiency in mild congestion while preventing unnecessary overhead.

\subsection{Experiment results with multiple congestion trees}
\label{sec:4-cong-tree}

\begin{figure*}[!htb]
    \centering
    \begin{subfigure}[b]{\linewidth}
      \centering
        \includegraphics[width=0.6\linewidth]{figures/experiments/legend.pdf}
        \end{subfigure}
        \hfill
      \begin{subfigure}[b]{\linewidth}
      \centering
        \includegraphics[width=\linewidth]{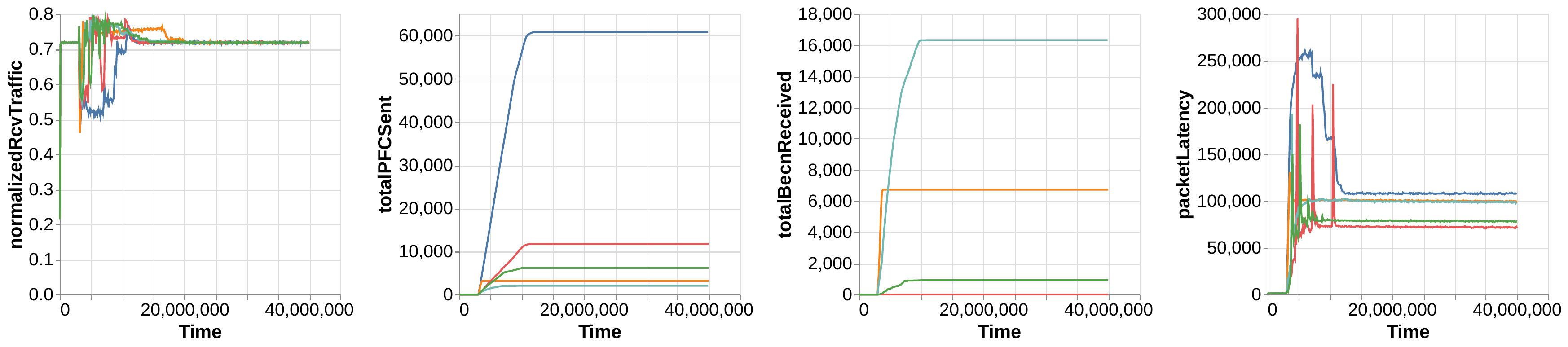}
        \end{subfigure}
      \caption{4HS Traffic pattern 64 MIN network.}
        \label{fig:2-cong-becn}
      
\end{figure*}

This scenario stresses ICI’s scalability by distributing congestion across four destinations, increasing the number of distinct congesting flows beyond the CFT capacity. As shown in Figure~\ref{fig:2-cong-becn}, standard ICI begins to degrade, while DCQCN maintains good throughput, but only with appropriately tuned thresholds.

However, when DCQCN thresholds are set too conservatively (high $K_{\min}$, late activation), it reacts too late, allowing buffer buildup. Conversely, overly aggressive thresholds cause overreaction to transient bursts. Tuning DCQCN optimally is challenging and workload-dependent.

ICI+ addresses this by integrating ICI’s flow-level visibility with adaptive ECN triggering. When ICI reaches its capacity, ICI+ seamlessly offloads congestion signaling to DCQCN, but with context-aware thresholds. As a result, ICI+ outperforms both standalone DCQCN and the uncoordinated ICI-DCQCN combination.

Notably, ICI+ reduces the peak number of generated BECNs by up to $\times 32$ compared to baseline DCQCN (Figure~\ref{fig:2-cong-becn}), significantly lowering control-plane overhead and avoiding network-wide oscillations.

\subsection{Google-All traffic and one congestion tree (long flow)}
\label{sec:google-1tree}

This experiment combines realistic short-flow traffic with a sustained incast. The simulation ends upon delivery of all 2 million \texttt{Google\_All} messages, so time axes vary across runs.

\begin{figure*}[!htb]
    \centering
    \begin{subfigure}[b]{\linewidth}
      \centering
        \includegraphics[width=0.6\linewidth]{figures/experiments/legend.pdf}
        \end{subfigure}
      \begin{subfigure}[b]{\linewidth}
      \centering
        \includegraphics[width=\linewidth]{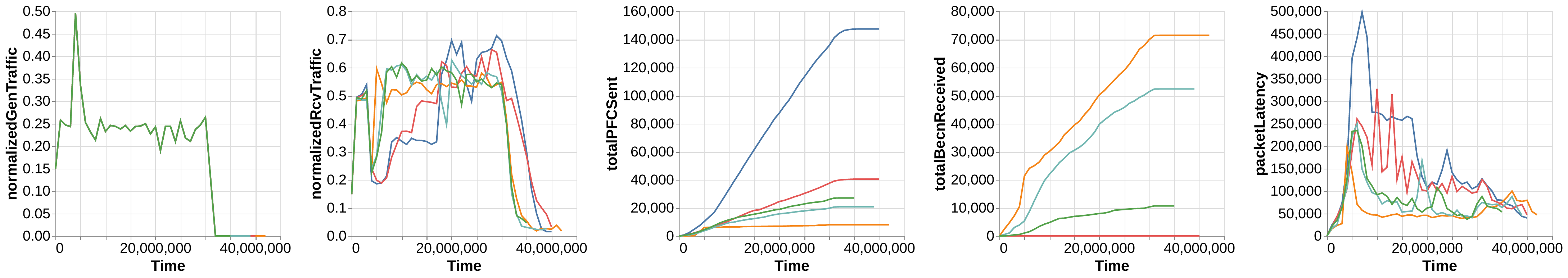}
        
      \end{subfigure}
      \hfill
      \caption{Google All Traffic pattern + Long Flow 64 MIN network.}
        \label{fig:google-becn}
\end{figure*}

Despite the differing durations, ICI+ consistently reduces the total number of BECNs generated (Figure~\ref{fig:google-becn}). More importantly, in terms of Flow Completion Time (FCT), ICI+ outperforms the best-performing baseline (ICI-DCQCN) across median and 95th percentile latencies.

\begin{figure}[!htb]
    \centering
    \includegraphics[width=\linewidth]{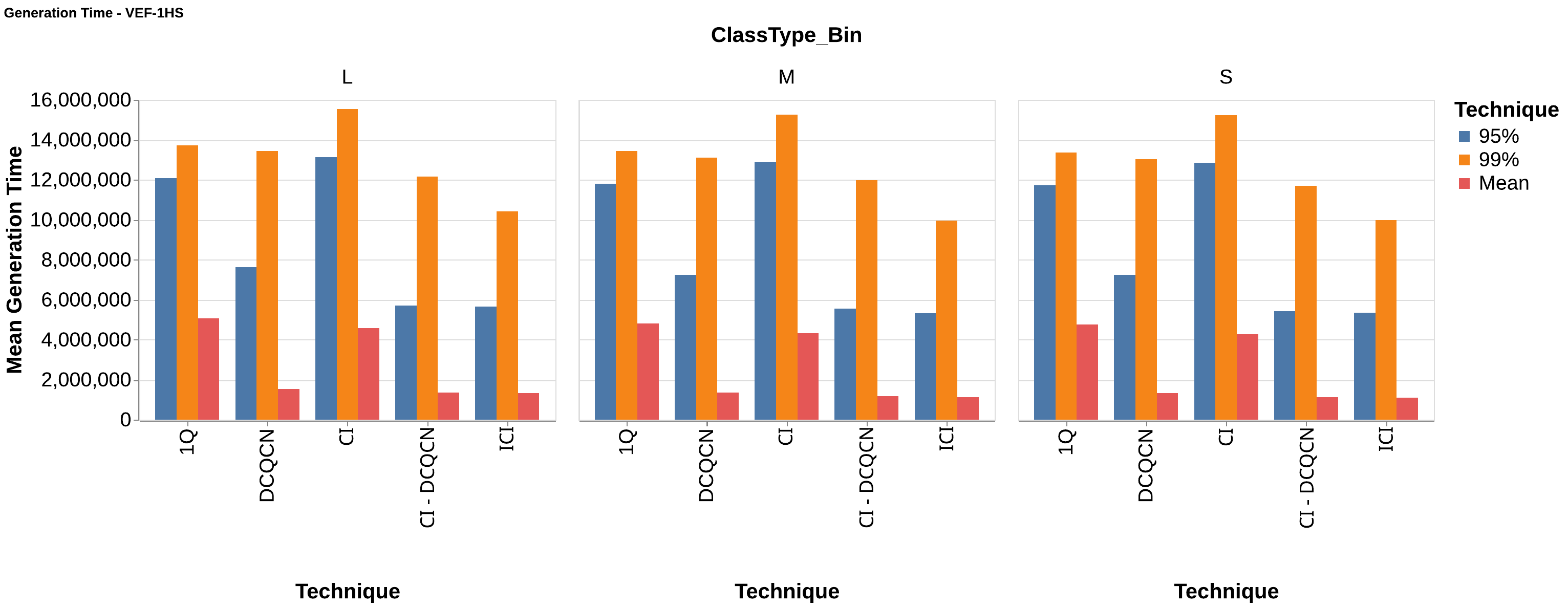}
    \caption{FCT Google traffic 64 MIN network}
    \label{fig:google-fct-long}
\end{figure}

When extended to longer-running versions of this workload (Figure~\ref{fig:google-fct-long}), ICI+ achieves the best tail latency (99th percentile), reducing FCT by up to 23\% compared to ICI and 31\% compared to DCQCN. This improvement stems from ICI+’s ability to suppress unnecessary congestion signals while maintaining rapid response when needed.

\subsection{Google-All traffic with four congestion trees}
\label{sec:google-4burst}

\begin{figure*}[!htb]
    \centering
    \begin{subfigure}[b]{\linewidth}
      \centering
        \includegraphics[width=0.6\linewidth]{figures/experiments/legend.pdf}
        \end{subfigure}
        \hfill
      \begin{subfigure}[b]{\linewidth}
        \centering
        \includegraphics[width=\linewidth]{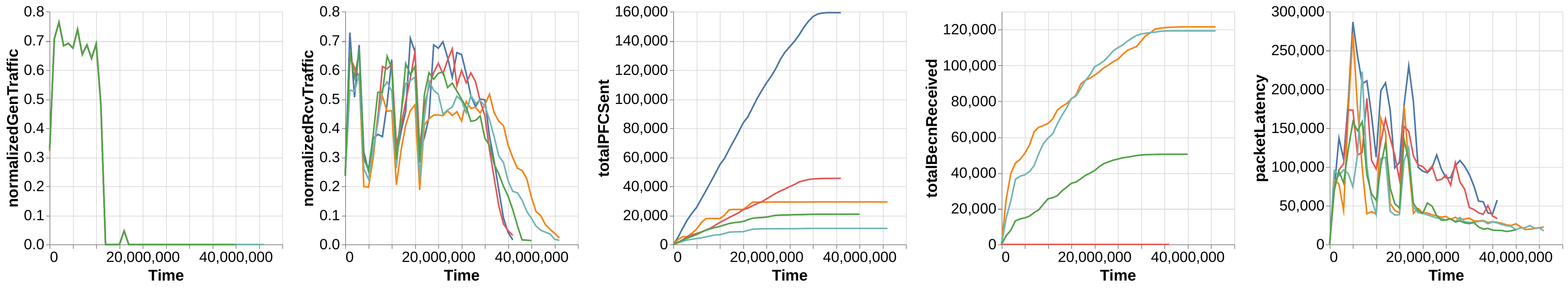}
        \end{subfigure}
      \caption{Google All Traffic pattern + 4 Congestion Burst 64 MIN network.}
        \label{fig:google-burst-throughput}
      
\end{figure*}

Here, the incast traffic is split into four small, frequent bursts. DCQCN struggles in this regime due to its latency in detecting micro-congestion and its tendency to overreact, leading to unnecessary rate reductions. As shown in Figure~\ref{fig:google-burst-throughput}, DCQCN degrades throughput more than no congestion control at all.

ICI performs well, as most bursts are absorbed without overflowing the CFT. However, ICI+ further improves efficiency by selectively engaging ECN marking only when bursts accumulate or persist, thus avoiding overreaction.

\begin{figure}[!htb]
    \centering
    \includegraphics[width=\linewidth]{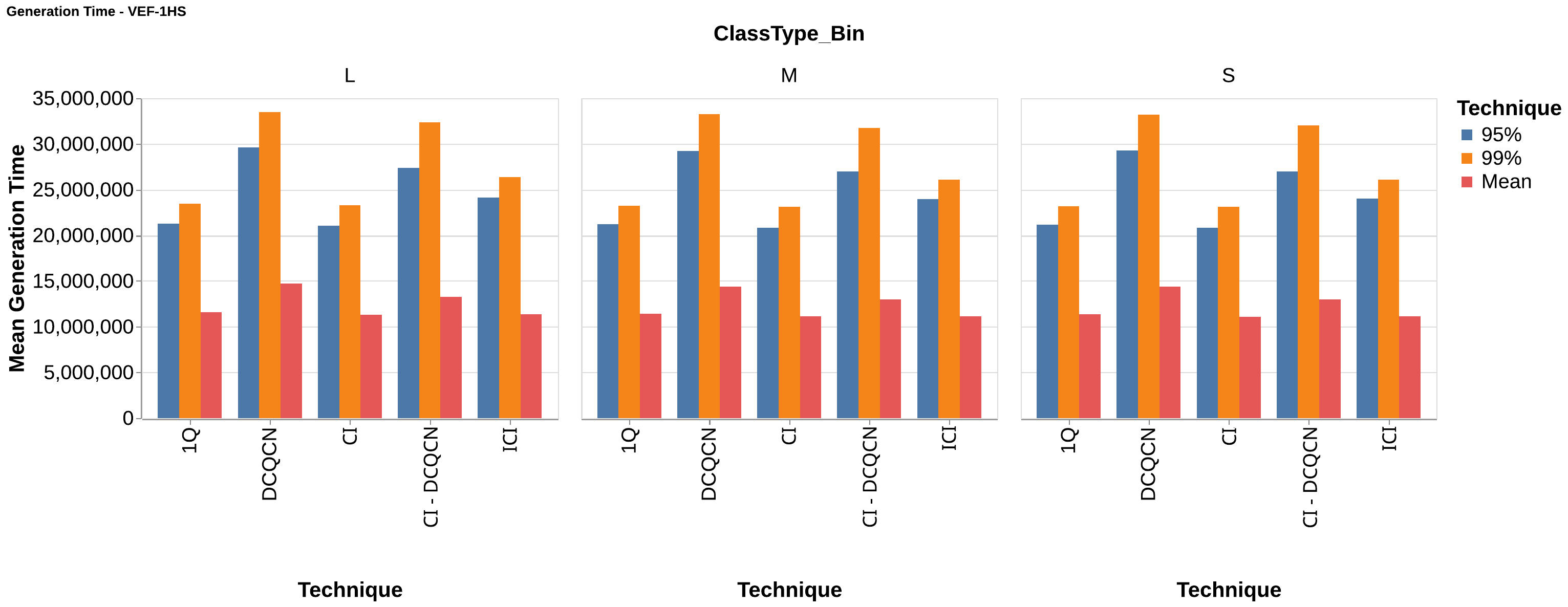}
    \caption{FCT Google All + 4 Congestion Burst 64 MIN network.}
    \label{fig:fct-4Burst}
\end{figure}

As a result, ICI+ achieves better FCT and throughput than both ICI and DCQCN, demonstrating its robustness in dynamic, bursty environments \ref{fig:fct-4Burst}.

\section{Conclusions and Future Work}
\label{sec:conclu}

This paper presents ICI+, a coordinated congestion control framework that enhances both ICI and DCQCN in modern RDMA-based data centers. ICI+ introduces a hybrid decision engine that dynamically selects or combines congestion responses based on flow-level visibility and buffer pressure, avoiding redundant or conflicting actions.

Through extensive evaluation across synthetic and real-world workloads, we show that ICI+:
\begin{itemize}

\item Reduces BECN generation by up to $\times 32$,
    \item Improves tail latency (99th percentile) by up to 31\%,
    \item Maintains high throughput under bursty and multi-congestion scenarios,
    \item Avoids overreaction in microburst conditions.
\end{itemize}

\section{Acknowledgements}
This research publication is part of the TED2021-130233B-C31 project, funded by MCIN/AEI/10.13039/501100011033.

\appendix

\bibliographystyle{elsarticle-num} 
\bibliography{database.bib}

\end{document}